\documentstyle[12pt,axodraw]{article}
\def\be{\begin{equation}}
\def\ee{\end{equation}}
\def\bea{\begin{eqnarray}}
\def\eea{\end{eqnarray}}
\def\q{{\bf q}}
\def\p{{\bf p}}
\def\k{{\bf k}}

\def\Pone{{\bf P}_1}
\def\Ptwo{{\bf P}_2}
\def\lam{\lambda}
\def\Lam{\Lambda}
\def\Gam{\Gamma}
\def\gam{\gamma}
\def\ie{{\it i.e.\ }}
\def\eg{{\it e.g.\ }}

\begin{document}
\begin{titlepage}
\begin{flushright}
SHEP 99-06 \\
hep-th/9906166
\end{flushright}
\vspace{.4in}
\begin{center}
{\Large{\bf Convergence of derivative expansions of the
renormalization group}}
\bigskip \\ Tim R. Morris and John F. Tighe \\
\vspace{\baselineskip}
{\it Department of Physics, University of Southampton, Highfield}\\
{\it Southampton SO17 1BJ, UK}

\mbox{} \\
\vspace{.5in}
{\bf Abstract}\bigskip \end{center} \setcounter{page}{0}
We investigate the convergence of the derivative expansion of the
exact renormalization group,
by using it to compute  the $\beta$
function of scalar $\lam\varphi^4$ theory.
We show that the derivative expansion of the Polchinski flow equation
converges at one loop for certain fast falling smooth cutoffs.
The derivative expansion of the Legendre flow equation trivially converges 
at one loop, but also at two loops: slowly with 
sharp cutoff (as a momentum-scale expansion), and rapidly in the case of
a smooth exponential cutoff. Finally, we show that the two loop contributions
to certain higher derivative 
operators (not involved in $\beta$) have divergent 
momentum-scale expansions for sharp cutoff, but 
the smooth exponential cutoff
gives convergent derivative expansions  
for all such operators with any number of derivatives.
\end{titlepage}

\section{Introduction and discussion}

It is obvious that there is a need to have good analytic
non-perturbative approximation methods for quantum field theory.
The exact renormalization group provides a powerful framework
for formulating such approximations because the calculations can
be phrased directly in terms of renormalized (\ie continuum)
quantities, thus trivially 
preserving a crucial property of quantum field
theory, namely renormalizability (equivalently the existence
of a continuum limit) \cite{morris1,morris4}.
In recent years it has been established that the derivative expansion
of the effective action, taken to some finite order, provides a 
robust and accurate non-perturbative
approximation for scalar field theory \cite{morris1}--\cite{WS}.

Whilst the validity of this statement currently rests mainly on empirical
fact, 
it is proven that the derivative expansion behaves correctly in various 
limits. We know that the lowest order
of the derivative expansion (the local potential approximation \cite{lpa})
is in a sense exact in the large $N$ limit  
(of \eg  the $N$-vector model) \cite{largens,ln2,on}. 
The form of the large field behaviour is also correctly reproduced by the 
expansion \cite{morris4,morris3,mass}.

Of course it is a challenging task to prove the applicability
of the derivative expansion non-perturbatively and in all generality.
Note in particular that this is {\sl not} a controlled
expansion in some small parameter. The approximation lies in neglecting
higher powers of $p/\Lambda$, where $\Lambda$ is the effective cutoff
and $p$ some typical momentum. But in the flow equations, the typical
momenta that contribute are themselves of order $\Lambda$. 
Thus the expansion is a numerical one, and an important but difficult
question to answer is whether this numerical series converges,
and indeed if so, whether it converges to the exact value.

Here we establish and extend some results on the weak coupling regime,
first announced in \cite{morris5}. In particular we show that
for the Legendre flow equation and an exponential cutoff, the derivative
expansion computation for the $\beta$ function of
$\lam{\varphi}^4$ theory\footnote{in four dimensions with
$\varphi\leftrightarrow-\varphi$ symmetry} converges at one and two loops 
to the correct result. 

We see then that the derivative
expansion interpolates between exact results
in a number of different limits, and in particular must be accurate
even at only-moderately weak coupling. Viewed from this perspective,
it is hardly surprising that the fully
non-perturbative results tend to be so accurate \cite{morris1}--\cite{WS}.

On the basis of intuitive arguments and preliminary calculations,
one of us suggested that
the momentum expansion of the Legendre flow equation with sharp cutoff,
should converge most rapidly \cite{morris1,morris2}. In fact, while
for the Legendre equations
all cutoffs trivially supply the exact answer for the $\beta$ 
function at one loop, the two-loop sharp cutoff
result converges only very slowly to the exact answer. Utilising
these results we can see that the derivative expansion result
for the two-loop contribution to the four-point vertex, but expanded
to second or higher order in external momenta, in fact
fails to 
converge for sharp cutoff. Actually these results again concur with the `phenomenology',
since non-perturbatively, results with sharp cutoff are
consistently worse than smooth cutoff versions \cite{mass,zak}.

A similar investigation for smooth cutoff shows that the 
derivative expansion series converges at two loops, no matter how
high order the expansion in external momenta is taken.

The derivative expansion of the Polchinski flow equation fails to
converge even at one loop for sharp-cutoff, power-law cutoff,
exponential cutoff and many other similar forms. Remarkably however, we
identify some forms of steeply falling cutoff that do result in
convergence.  Their existence underpins the competitive results that
may be obtained with this method to $O(\partial^2)$ 
\cite{Ball}--\cite{polder}.
(These results tend to be obtained for all cutoffs, and depend on a 
finite number of parameters: none for $O(\partial^0)$, three for 
$O(\partial^2)$ \cite{Ball}\cite{morris4,morris3,morris5,threeps}.)

Momentum scale expansions with sharp cutoff and 
derivative expansions with power-law cutoff, applied to the Legendre
flow equations, share 
the property that reparametrization invariance is preserved 
\cite{morris3,morris5}.
However, we will show that derivative expansions with power law
cutoff actually fail badly at two loops, in the sense that at some order
greater than $O(\partial^4)$
(which we determine) the result is infinite because the
momentum integrals fail to converge. The reason is that high order 
derivative expansions with power-law cutoff result in contributions that 
actually diverge for small coupling and therefore fail even qualitatively
in this regime. These conclusions are puzzling in view of the
very competitive non-perturbative $O(\partial^2)$
results obtained with power-law 
coupling \cite{on}--\cite{morris5,zak,twod}.\footnote{However, exact
$O(\partial^2)$ results with exponential coupling do appear to be better 
still \cite{Christoph}.} 

Of course it must be borne in mind that
there is no real conflict here: it is quite possible that these
$O(\partial^2)$ approximations provide accurate models for scalar field
theory, especially since they interpolate between exact results in
various limits, even if the higher orders of the derivative expansion
do not result in convergent
series. It is rather that convergence of the derivative expansion at low
orders of the coupling {\sl guarantees} the accuracy of high orders
of the derivative expansion in this regime.

All the calculations presented in this
paper have been performed directly in terms of renormalized quantities,
which is one of the beauties of this approach \cite{morris1,morris4,bonini},
but  we have explicitly checked that the same results are
obtained with the `traditional' approach of introducing an overall
cutoff $\Lam_0$ and allowing this to tend to infinity.

There are many possible extensions of this work.  
These are discussed at the end of the paper. The plan
for the rest of the paper is as follows. In section 2 we consider the
Wilson/Polchinski flow equation, demonstrate that the correct
$\beta$ function appears at one loop for any cutoff when solved
exactly, and then analyse the numerical series that follows from the
derivative expansion. Apart from the summary, the remaining sections
are concerned with the Legendre flow equations.
Section 3 treats one loop order.  The derivative expansion has
no effect in this case and the correct $\beta$ function is always
obtained. In section 4 we discuss the specific case of the
sharp cutoff Legendre flow equation, calculating the 
two-loop $\beta$ function as a convergent momentum-scale expansion.
In this section we also identify the momentum-squared operator whose
series fails to converge. Section 5 treats the case of a
smooth exponential cutoff, demonstrating that the resulting numerical
series for the two-loop $\beta$ function converges 
rapidly to the correct result.
This time considering higher powers of external momenta does not alter
the conclusions: all these operators have convergent series.
Finally, in section 6 we show how the power law cutoff fails
at two loops. Section 7 contains a summary and
a discussion of future directions.

\section{Wilson/Polchinski flow equation} 

As Wilson showed \cite{wilson} the renormalization of quantum field
theories can be understood within the context of the flow of an
effective action $S{_\Lam}[\varphi]$ with an effective ultra-violet
(UV) cutoff $\Lam$.  Polchinski's version \cite{pol} of the flow equation
can be obtained from that of Wilson by substitution \cite{morris3}.

Firstly, let us define a modified propagator $\Delta_{UV}$ such that
$\Delta_{UV}={{C_{UV}(q^2/\Lam^2)}/{q^2}}$.  $C_{UV}(x)$ is an as
yet unspecified function\footnote{$x$ here stands for
$q^2/\Lam^2$. $C_{UV}$ is a function of only this, by 
Lorentz invariance and dimensions.} 
which acts as an UV cutoff and hence has the
properties $C_{UV}(0)=1$ and $C_{UV}\rightarrow0$ (sufficiently
fast) as $q\rightarrow\infty$. Similarly we can define
$\Delta_{IR}=C_{IR}(q^2/\Lam^2)/q^2$, where $C_{IR}=1-C_{UV}$
behaves as an infrared cutoff. Following
ref.\cite{morris1} we define  $K_\Lam={d\over{d\Lam}}\Delta_{UV}$ and
expand the flow equation for the $n$ point function, the
$\phi^n$ vertex of the effective action with $n$ external momenta:  
\bea\label{wpfl}
\lefteqn{{\partial\over\partial\Lam}S(\p_1,\cdots,\p_n;\Lam)=
\sum_{\left\{I_1,I_2\right\}}
S(-\Pone,I_1;\Lam)K_{\Lam}(P_1)S(\Pone,I_2;\Lam)} \nonumber \\
& & \hspace{2in} -{1\over 2}\int\!\!{d^4q\over(2\pi)^4}K_{\Lam}(q)
S(\q,-\q,\p_1,\cdots,\p_n;\Lam),
\eea
where $I_1$ and $I_2$ are disjoint subsets of external momenta such
that $I_1\cap I_2=\emptyset$ and $I_1\cup
I_2=\{\p_1,\cdots,\p_n\}$. The sum over ${\left\{I_1,I_2\right\}}$
utilises the Bose symmetry so pairs are counted only once \ie
${\left\{I_1,I_2\right\}}={\left\{I_2,I_1\right\}}$. The momentum
$\Pone$ is defined to be $\Pone={\sum_{\p_i\epsilon I_1}}\p_i$. 

We first consider the four point vertex exactly (\ie without a
derivative expansion) 
and obtain
the exact $\beta$ function. We define the renormalized
coupling $\lambda$ to be the four point vertex at zero momentum.
The only contribution comes from the tree level six point
function with two legs tied together to give the diagram 
in fig~\ref{figone}.

\begin{center}
\SetScale{1.0}
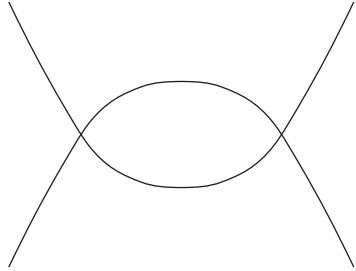
\begin{figure}[tbh]
\begin{picture}(220,100)(-70,0)
\Curve{(70,100)(80,80)(90,62)(100,46)(120,32)(134,30)
(136,30)(150,32)(170,46)(180,62)(190,80)(200,100)}
\Curve{(70,0)(80,20)(90,38)(100,54)(120,68)(134,70)
(136,70)(150,68)(170,54)(180,38)(190,20)(200,0)}
\end{picture}
\caption{Feynman diagram contributing to four point function 
at one loop}
\label{figone}
\end{figure}
\end{center}

The tree level six point vertex follows from
integrating up (\ref{wpfl}) with $\lambda$ as the four-vertex.
Substituting back into (\ref{wpfl}) we thus have (compare 
\cite{morris1,bonini})
\bea 
\lefteqn{{\partial\over\partial\Lam}\lam=
3{\lam}^2\int\!\!{d^4q\over(2\pi)^4}K_{\Lam}(q)
\left[\int_{\Lam}^{\infty}{d\Lam_1}K_{\Lam_1}(q)\right]} \label{wp6pt}
\\
& & =
{3\lam^2}\int\!\!{d^4q\over(2\pi)^4}K_{\Lam}(q)
\Delta_{IR}(q^2/\Lambda^2) \label{wptder}
\\
& & =
{6{\lam}^2\over(4\pi)^2}{1\over\Lambda}\int^\infty_0\!\!dx\, 
 C_{IR}'(x) C_{IR}(x) \label{wpots} 
\\
& & = 
{3{\lam}^2\over (4\pi)^2} {1\over \Lam}
\left[C_{IR}^2(\infty)-C_{IR}^2(0)\right], \label{wpans}
\eea
where in (\ref{wpots}) the prime means differentiation with respect
to $x$
and in (\ref{wp6pt})
the tree-level six-point function provides the $\Lambda_1$ integral. 
Since the $\Lambda_1$ integral is UV convergent
we do not need an overall cutoff ($\Lambda_0$),  but instead go direct
to the continuum limit ($\Lambda_0=\infty$). 
Using the definition of $C_{IR}=1-C_{UV}$, we see that the $\beta$ function  
($\beta(\Lam)\equiv\Lam{\partial\over\partial\Lam}\lam$) is correctly 
\cite{zinn} reproduced as ${3{\lam}^2\over (4\pi)^2}$ at one loop. 
(We remind the reader that the one and two loop $\beta$ function
coefficients are universal for massless four dimensional $\lambda\phi^4$
theory.)

Now consider evaluating 
the effective action in terms of a derivative expansion.
This means that the $\Delta_{IR}$ in (\ref{wptder}), which is
the momentum dependent part of the classical six-point contribution,
must be expanded in $q$. The intermediate result (\ref{wpots}) is
thus replaced by
\be
\beta={6\lambda^2\over(4\pi)^2}\sum_{n=1}^\infty{C_{UV}^{(n)}(0)\over n!}
\int_0^\infty\!\!\!\!\!\!dx\, x^n C'_{UV}(x), \label{watsit}
\ee
where $C_{UV}^{(n)}$ is the $n^{\rm th}$ $x$-derivative of $C_{UV}(x)$.
Taking the specific example of sharp cutoff $C_{UV}=\theta(1-x)$
we trivially obtain convergence to the wrong answer, namely zero, 
since $C_{UV}^{(n)}(0)=0$
for all $n\ge1$. However for sharp cutoff, the derivative expansion 
actually does not exist, and a careful determination of the
sharp cutoff limit leads one to the Legendre equations even in the exact
case \cite{morris2}. Therefore we will consider the sharp cutoff from now
on only in the Legendre flow equations. If we choose a cutoff that
decays as a power of $q^2$, it is clear that at some $n$ the integral
in (\ref{watsit}) diverges. In order to avoid this problem we must
make the cutoff decay faster than a power.
If we choose an exponential
cutoff of the form $C_{UV}={\rm e}^{-q^2/\Lam^2}$ however, we still have
problems:
\be
\beta={6{\lam}^2\over (4\pi)^2}
\sum_{n=1}^{\infty}(-1)^{n+1}. \label{wpnocon}
\ee
Clearly this does not converge. In fact, the reader can readily confirm
that exponential decay with any power ($C_{UV}={\rm e}^{-x^m}$) still fails
to give a convergent derivative expansion. The $\sim1/n!$ behaviour
of the Taylor expansion coefficients is not enough in these cases
to overcome the $\sim n!$ behaviour of the integrals. Clearly any
cutoff that behaves as $\sim {\rm e}^{-x^m}$ for large $x$ will thus also
fail. Indeed if these cutoffs are not entire as functions of complex $x$,
the Taylor expansion coefficients will decay only as a power $\sim 1/R^n$
($R$ the distance from the origin to the nearest singularity)
and will result in badly divergent numerical series.

However, we can obtain convergence 
if we insist that the UV cutoff is entire
but falls fast enough as $x\to\infty$ to ensure that the integrals in
(\ref{watsit}) grow much slower than n! {}\footnote{In the limit of sharp
cutoff the integrals are all 1.}  With 
$C_{UV}(x)=\exp\left(1-{\rm e}^x\right)$ for example,
we obtain
\be
\beta={3\lambda^2\over(4\pi)^2}\left\{1.193+0-0.194-0.060+0.032
+\cdots\right\},
\ee
with the series in braces
summing to 1.000286 after twelve terms. With $C_{UV}=
\exp\left\{{\rm e}-\exp\left({\rm e}^x\right)\right\}$ we obtain
\be
\beta={3\lambda^2\over(4\pi)^2}\left\{1.278-0.164-0.130-0.014+0.019
+\cdots\right\},
\ee
which also converges,
with the series in braces summing to 0.999551 after twelve terms. (The
integrals of (\ref{watsit}) were calculated numerically.) 

It would be very interesting to investigate if these forms of cutoff
yield convergent derivative expansions at higher loops also. However,
in this paper we now turn to the Legendre flow equations: these have
inherently better convergence properties
at the very least because, being 1PI (one particle
irreducible), there are no tree-level corrections and thus numerical
series arising from derivative expansion do not appear until the
two loop level.

\section{Legendre flow equation at one loop}

Following the notation of ref.\cite{morris1},
see also \cite{leg,leg2}, 
we have the Legendre flow equation for a
general cutoff (with the vacuum energy dropped):
\be\label{legfl}
{\partial\over\partial\Lam}\Gam[\varphi^c]=-{1\over 2}{\bf{tr}}
\left\{{K_{\Lam}\over
(1+\Delta_{IR}\Sigma)^2}.\hat{\Gam}.(1+[\Delta_{IR}^{-1} +
\Sigma]^{-1}.\hat{\Gam})^{-1}\right\}, 
\ee 
where $\Gam[\varphi^c]$ is the generator of 1PI Greens functions,
$\varphi^c$ is the classical field, $\hat{\Gam}[\varphi^c]$ 
the second field differential 
less $\Sigma$, the field independent part \ie
the effective self energy. Eqn (\ref{legfl}) can
be expanded in terms of $\varphi^c$ to give the expanded Legendre flow
equation although care is required when taking the sharp cutoff
limit \cite{morris1,morris2}. 
However at one loop (irrespective of exact form of cutoff), the
only contribution to the flow of the four point function will arise
from 
\bea\label{legfl1lp}
\lefteqn{{\partial\over\partial\Lam} \Gam(\p_1,\p_2,\p_3,\p_4;\Lam)
= \int\!\!{d^4q\over(2\pi)^4}K_{\Lam}(q)}
\nonumber \\
& & \times\sum_{\left\{I_1,I_2\right\}}\Gam(\q,-\q-\Pone,I_1;\Lam)
\Delta_{IR}(|\q+\Pone|)\Gam(\q-\Ptwo,-\q,I_2;\Lam),
\eea
where the notation is the same as used in (\ref{wpfl}). With $\lambda$
defined as the four-point 1PI vertex at zero momenta 
and substituting $\lam$ --the tree-level result, in the right hand side,
this reduces to 
\be\label{lg1lpdex}
{\partial\over\partial\Lam} \lam = 3\lam^2\int\!\!{d^4q\over(2\pi)^4} 
{1\over q^4} \left({d\over{d\Lam}} C_{IR}(q^2/\Lam^2)\right) 
C_{IR}(q^2/\Lam^2)
\ee
Derivative expansion corresponds to an expansion in the external
momenta of $\Gamma$'s vertices, but in this case the relevant vertices
have no external momentum dependence.
It is easy to see that (\ref{lg1lpdex}) is the same as
(\ref{wptder}) and thus for the Legendre equations, even
in the derivative expansion, we obtain the
exact one-loop $\beta$ function ${3{\lam}^2\over
(4\pi)^2}$ irrespective of the exact form of cutoff function.

\section{Sharp cutoff at two loops}
 
With due caution the sharp cutoff limit [$C_{UV}=\theta(\Lam-q)$] of
(\ref{legfl}) is taken \cite{morris1}, and after expanding in
$\varphi^c$ we obtain the sharp cutoff expanded Legendre flow equation
for $n$ external momenta: 
\be\label{legflsh}
{\partial\over\partial\Lam} \Gam(\p_1,\cdots,\p_n;\Lam)
= \int\!\!{d^4q\over(2\pi)^4}{{\delta(q-\Lam)}\over
{q^2+\Sigma(q;\Lam)}}E(\q,\p_1,\cdots,\p_n;\Lam),
\ee
where
\bea\label{legflE}
\lefteqn{E(\q,\p_1,\cdots,\p_n;\Lam) = - {1\over 2} 
\Gam(\q,-\q,\p_1,\cdots,\p_n;\Lam)} \nonumber \\ 
& &  \hspace{0.1in}+\sum_{\left\{I_1,I_2\right\}}\Gam(\q,-\q-\Pone,I_1;\Lam)
G(|\q+\Pone|;\Lam)\Gam(\q-\Ptwo,-\q,I_2;\Lam) \nonumber\\ 
& &  -\sum_{\left\{I_1,I_2\right\},I_3}\Gam(\q,-\q-\Pone,I_1;\Lam)
G(|\q+\Pone|;\Lam) \times \nonumber\\ 
& & \hspace{0.4in} \Gam(\q+\Pone,-\q+\Ptwo,I_3;\Lam)
G(|\q-\Ptwo|;\Lam)\Gam(\q-\Ptwo,-\q,I_2;\Lam) \nonumber \\
& & \hspace{4in}+\cdots \hspace{0.1in}.
\eea
Similarly to before, ${\bf P}_i=\sum_{\p_j\in I_i}\p_j$ and
$\sum_{\{I_1,I_2\},I_3,\cdots,I_m}$ is a sum over disjoint subsets
$I_i\cap I_j=\emptyset$ $(\forall i,j)$ with
${\bigcup_{i=1}^{m}}I_i=\{\p_1,\cdots,\p_n\}$.  Again, the
symmetrization $\{I_1,I_2\}$ means this pair is counted only
once. $G(p;\Lam)$ is defined by
\be\label{Gsh}
G(p;\Lam)\equiv
{\theta(p-\Lam)\over{p^2+\Sigma(p;\Lam)}}, \nonumber
\ee
where $\Sigma$ is again the (field independent) self energy.

By iteration we can now solve (\ref{legflsh}) to two loop
order.  We split the four point function into
two parts, momentum free [$\lam(\Lam)$] and momentum dependent \cite{bonini}:
\bea\label{ren4pt}
\Gam(\p_1,\p_2,\p_3,\p_4;\Lam) &=& \lam(\Lam) +
\gam(\p_1,\p_2,\p_3,\p_4;\Lam), \\
{\rm where}\hskip2cm\gam(0,0,0,0;\Lam) &=&0. \label{ga0}
\eea
Therefore at one loop
\bea\label{shfl1lp}
\lefteqn{\gam(\p_1,\p_2,\p_3,\p_4;\Lam)}  \nonumber \\
& & = - \lam^2 \int_{\Lam}^{\infty}d\Lam_1
\int\!\!{d^4q\over(2\pi)^4}{{\delta(q-\Lam_1)}\over q^2}\sum_{i=2}^{4} 
\left\{{\theta(|\q+{\bf\cal P}_i|-\Lam_1)
\over (\q+{\bf\cal P}_i)^2}-{\theta(q-\Lam_1)\over q^2} 
\right\} \label{sh1lp} \\
& & =  -{\lam^2\over4\pi^3}\sum_{i=2}^4\int_\Lam^\infty{dq\over q}
\left\{
-{1\over2}+\int^1_{-1}\!\!dx\ \theta(2x+{\cal P}_i/q)\,
{\sqrt{1-x^2}\over 
1+2x{\cal P}_i/q+ {\cal P}_i^2/q^2} \right\}
\label{sh1lpex} \\
& & = +{\lam^2\over4\pi^3}\sum_{i=2}^{4} 
\left\{{1\over6}{{\cal P}_i\over \Lam}
+{1\over720}\left({{\cal P}_i\over\Lam}\right)^3+{3\over44800}\left({{\cal
P}_i\over \Lam}\right)^5
+\cdots \right\}, \label{sh1lpans}
\eea
where ${\bf{\cal P}_i}=\p_1+\p_i$ and $x={\bf{\cal P}_i}\cdot\q/{\cal P}_i$.
Note that the subtraction of the momentum independent part in 
(\ref{sh1lp}) ensures that the $\Lam_1$ integral converges,
allowing (again) the upper limit to be set as $\infty$. 
In (\ref{sh1lpex}) we perform the $\Lam_1$ integral, noting that
effectively here $\theta(0)={1\over2}$ \cite{morris1}. 
By absorbing the step function into the $x$ limit, the term in braces
may be expanded in momentum-scale 
${\cal P}_i=|{\bf{\cal P}_i}|$ \cite{morris1,morris2}
to give (\ref{sh1lpans}). Alternatively the step function may be
expanded directly \cite{morris1}  
\bea
\theta\left({{\cal P}_i\over 2q} + x\right)=\theta(x)
+\sum_{n=1}^{\infty}{1\over {n!}}\left({\cal P}_i\over2q\right)^n
\delta^{(n-1)}(x),
\eea
where $\delta^{(n-1)}(x)$ is the $(n-1)$th derivative of $\delta(x)$
with respect to x.  Of course (\ref{sh1lpans})
agrees with the bare version already computed in ref.\cite{morris1}.

We will also need the renormalized one-loop self-energy which we obtain
directly from the flow equation as follows:
\bea\label{seshdif}
\lefteqn{{\partial\over\partial\Lam}\Sigma(p;\Lam) = - {\lam\over
2}\int\!\!{d^4q\over(2\pi)^4}
{\delta(q-\Lam)\over q^2}}  \nonumber \\
& & \hspace{0.5in}= -{\lam\over (4\pi)^2}\Lam. 
\eea
Integrating up (\ref{seshdif}) we must not introduce a mass scale as
this is a massless theory.  Consequently, the uniquely determined self
energy has to be
\be\label{sesh}
\Sigma(p;\Lam) = -{\lam\over (4\pi)^2}{\Lam^2\over 2}
\ee
(with or without momentum-scale expansion).

At two loops, the diagrams in  (b) and (c) of fig~\ref{figtwo} will
contribute to the $\beta$ function.
Diagram (a) might be expected to be included, however on setting external 
momenta to zero the iterand of this topology vanishes by (\ref{ga0}). The
reason is that in
calculating renormalized quantities directly, (a) is already incorporated
in the one-loop running
$\lambda(\Lambda)$. (In the more traditional calculation, (a) only
has a divergent part which is then removed on renormalization.)

\begin{center}
\SetScale{0.7}
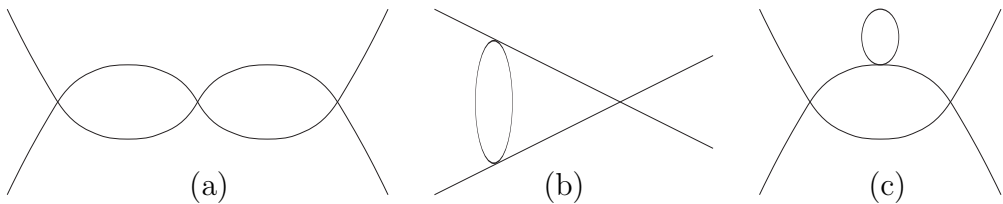
\begin{figure}[tbh]
\begin{picture}(410,100)(20,0)
\Curve{(70,100)(80,80)(90,62)(100,46)(120,32)(134,30)
(136,30)(150,32)(170,46)(175,54)(195,68)(209,70)
(211,70)(225,68)(245,54)(255,38)(265,20)(275,0)}
\Curve{(70,0)(80,20)(90,38)(100,54)(120,68)(134,70)
(136,70)(150,68)(170,54)(175,46)(195,32)(209,30)
(211,30)(225,32)(245,46)(255,62)(265,80)(275,100)}
\put(118,0) {(a)}
\Line(300,0)(450,75)
\Line(300,100)(450,25)
\Oval(332,50)(33,10)(0)
\put(252,0) {(b)}
\Curve{(475,100)(485,80)(495,62)(505,46)(525,32)(539,30)
(541,30)(555,32)(575,46)(585,62)(595,80)(605,100)}
\Curve{(475,0)(485,20)(495,38)(505,54)(525,68)(539,70)
(541,70)(555,68)(575,54)(585,38)(595,20)(605,0)}
\Oval(540,85)(15,10)(0)
\put(375,0) {(c)}
\end{picture}
\caption{Feynman diagrams contributing to the four point function 
at two loops.}
\label{figtwo}
\end{figure}
\end{center}

\vfill\eject

The flow equation at this order is
\bea\label{shmain2lp}
\lefteqn{{\partial\over\partial\Lam}\lam(\Lam) = {1\over \Lam}
{{3{\lam}^2}\over (4\pi)^2}} \nonumber\\
& & + \, 3{\lam}^3
\int\!\!{d^4q\over(2\pi)^4}{{\delta(q-\Lam)}\over q^2}  
\left[ {\Lam^2\over (4\pi)^2} {{\theta(q-\Lam)}\over q^4}
\right. \nonumber \\
& & \hspace{0.75in} - {1\over2}\int_{\Lam}^{\infty}\!\!d\Lam_1
\int\!\!{d^4p\over(2\pi)^4}{{\delta(p-\Lam_1)}\over p^2} 
\left\{ {3\theta^2(p-\Lam_1)\over p^4} \right. \nonumber \\  
& &  \hspace{1in} + {4\,\theta^2(|\p+\q|-\Lam_1)\over {|\p+\q|^4}} + 
{8\,\theta(|\p+\q|-\Lam_1)\,\theta(p-\Lam_1)\over {p^2|\p+\q|^2}} 
\nonumber\\
& & \hspace{1in} \left.\left.
+{4\theta(q-\Lam)\over q^2}\left( {\theta(|\p+\q|-\Lam_1)\over |\p+\q|^2}
- {\theta(p-\Lam_1)\over p^2}\right)
\right\}\right] .
\eea
In here, 
the first $O({\lam}^3)$ term arises from the expansion of 
the self energy (\ref{sesh}), \ie topology fig~\ref{figtwo}(c).
The next also gives rise to topology (c),
but is now generated by the one-loop six-point vertex
with two legs tied together. The six-point vertex also provides the
next two terms in the form of fig~\ref{figtwo}(b). The final line
arises from iterating the four-point function of (\ref{sh1lpans}) in
the flow equation. 

If we consider the first two (self energy) terms we find they evaluate
to $3{{\lam}^3\over (4\pi)^4}{1\over \Lam}$ and $-3{{\lam}^3\over
(4\pi)^4}{1\over \Lam}$ respectively ($\theta^2(0)\equiv1/3$ here 
\cite{morris1}), and thus cancel. Actually, it may be shown that these
self energy terms cancel for any cutoff.
Neither is affected by momentum-scale expansion, since 
no external momentum dependence is involved.
In the next two contributions the momentum expansion does affect the
embedded one-loop terms and corresponds in this case to an
expansion in $q/p$.
Proceeding as before,  we obtain the expansion of
the first of these to be
\be\label{sh6pt1}
-12{{{\lam}^3}\over (4\pi)^4}{1\over \Lam}{1\over
\pi}\left({\pi\over2}-{10\over9}+{\pi\over4}-{63\over
100}+{\pi\over6}-{7035\over 15680}+\cdots \right)
\ee
and of the other to be 
\be\label{sh6pt2}
-12{{{\lam}^3}\over (4\pi)^4}{1\over \Lam}{1\over
\pi}\left({\pi\over2}-{2\over9}-{1\over300}- {3\over15680}+\cdots
\right). 
\ee
We see that the second series converges rapidly, but the first series
is very slowly converging. We are sure that it
does actually converge because
we computed the first 80 terms. These continue
to oscillate and the partial sums show clear signs of slow convergence,
with 79 terms giving $S_{79}= .918$, and 80 terms $S_{80}=.879$
for the bracketed series 
in (\ref{sh6pt1}). The average of successive partial sums ${1\over2}
(S_{2n}+S_{2n-1})$ fits a form $a+b/n^2$ well for large $n$ and on
this basis we  estimate $S_\infty=.89828$ to 5dp.
The final part of (\ref{shmain2lp})
reproduces the previously published value \cite{morris1}  of 
\be
{{\lam}^3\over (4\pi)^4}{1\over \Lam} {1\over\pi}
\left(8+{1\over15}+{9\over2800}+\cdots \right).
\ee
At this level of perturbation theory wave function
renormalization appears through $\Sigma(k;\Lam)|_{O(k^2)}=[Z(\Lam)-1]k^2$
 arising from fig~\ref{figwfn2lp}, 
\bea
\lefteqn{ k^2{\partial\over\partial\Lam}Z(\Lam) } \nonumber
\\
& & \left. =\lam^2\int\!\!{d^4q\over(2\pi)^4}{
{\delta(q-\Lam)}\over q^2}
\int_{\Lam}^{\infty}d\Lam_1\int\!\!{d^4p\over(2\pi)^4} 
{\delta(p-\Lam_1)\over p^2} 
{\theta(|\p+\q+\k|-\Lam_1)\over |\p+\q+\k|^2}
\right|_{O(k^2)} \nonumber
\\
& & \left. =-{\lam^2\over 4\pi^3}\int\!\!{d^4q\over(2\pi)^4}{
{\delta(q-\Lam)}\over q^2}\left\{{1\over 6}{|\q+\k|\over
\Lam}+{1\over 720}{|\q+\k|^3\over \Lam^3}+{3\over 44800}{|\q+\k|^5\over
\Lam^5}+\cdots\right\} \right|_{O(k^2)} \nonumber
\\ 
& & = -{\lam^2k^2\over (4\pi)^4} {1\over \Lam} {1\over \pi}
\left({1\over 2}+{1\over 48}+{3\over 1280}+\cdots\right). \label{beesknees}
\eea
Note that the second line is the expanded one-loop four-point
vertex (\ref{sh1lpans}). The net effect of expanding to second order in
$k$ and averaging over the angles, is to convert $|\q+\k|^n$ into
${1\over8}n(n+2)q^{n-2}k^2$. 

\begin{center}
\SetScale{1.0}
\begin{figure}[tbh]
\begin{picture}(100,100)(0,0)
\Oval(200,50)(40,40)(0)
\Line(125,50)(275,50)
\end{picture}
\caption{Feynman diagram contributing to wave function renormalization
at two loops.}
\label{figwfn2lp}
\end{figure}
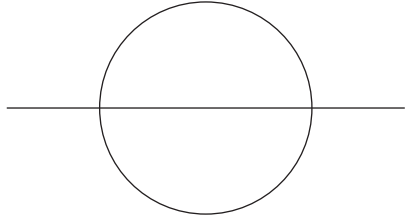
\end{center}
 
As a consequence, the properly normalised
definition of the $\beta$ function is now
$\beta(\Lam)=\Lam{\partial\over\partial\Lam}\left({\lam(\Lam)/
Z^2}\right)$. Collecting all terms, the $\beta$ function at two loops for the
Legendre flow equation with sharp cutoff is found to be 
\bea\label{betash}
\lefteqn{\beta(\Lam) = \Lam{\partial\over\partial\Lam}\lam -
2\lam\Lam{\partial\over\partial\Lam}Z(\Lam)} \nonumber 
\\
& & = 3{{\lam^2}\over {(4\pi)^2}} -{{{\lam}^3}\over
(4\pi)^4}{1\over\pi} \left\{
12\left({\pi\over2}-{10\over9}+{\pi\over4}-{63\over
100}+{\pi\over6}-{7035\over 15680}+\cdots \right) \right.  
\nonumber \\
& & \hspace{0.5in}  + 12\left({\pi\over2}-{2\over9}-
{1\over300}-{3\over15680}+\cdots \right) 
- \left(8+{1\over15}+{9\over2800}+\cdots
\right) 
\nonumber \\
& & \hspace{2.0in} \left. -\left(1+{1\over 24}+{3\over
640}+\cdots\right) 
\right\},
\eea
where in the first line we use $Z(\Lambda)=1+O(\lam^2)$.
Using our above result for the slow series, we see that the rest
rapidly converges towards the exact expression
\cite{zinn}
\be\label{truebeta}
\beta(\Lam)=3{{\lam^2}\over {(4\pi)^2}} - {17\over 3}{{\lam^3}\over 
{(4\pi)^4}}.
\ee 

Our observation below (\ref{beesknees}) however, shows that for
second order in the external momentum, equivalently
$O(\partial^2)$ operators, the $n^{\rm th}$ coefficient of momentum-scale
expansion  is effectively
multiplied by $\sim n^2$. For an $O(\partial^{2r})$ operator, 
the $q^n$ term is that
of the derivative-free operator (\ie with the same
number of fields) with $q^n$ replaced by 
$\sim\left(n/2\atop2r\right)q^{n-2r}k^{2r}$ 
which yields a multiplier $\sim n^{2r}$ for large
$n$. Unless the coefficients in
the numerical series fall faster than a power of $n$ (\eg
$\sim1/n!$ or $1/R^n$, $R>1$), we will be able to find a
sufficiently high derivative operator whose derivative expansion series
fails to converge. While (probably) all but one of the sharp-cutoff
series do have coefficients that fall faster than a power of $n$, the
slow series (\ref{sh6pt1}) barely manages to converge at all.  Thus it
is clear without calculation, that the contribution
fig~\ref{figtwo}(b) taken to second or higher
order in its external momenta \ie in particular the $O(k^{2r})$ $r\ge1$
coefficients of
\be
-6{\lam}^3
\int\!\!{d^4q\over(2\pi)^4}{{\delta(q-\Lam)}\over q^2}  
\int_{\Lam}^{\infty}\!\!d\Lam_1
\int\!\!{d^4p\over(2\pi)^4}{{\delta(p-\Lam_1)}\over p^2} 
 {\theta^2(|\p+\q+\k|-\Lam_1)\over {|\p+\q+\k|^4}},
\ee 
yield momentum-scale expansion series that fail to converge.

\section{Smooth cutoff}

We return to smooth cutoffs,
this time within the context of the Legendre flow equation. Eqn
(\ref{legfl}) can now be expanded in $\varphi^c$ directly without the
difficulties that arise from taking the sharp limit \cite{morris1}.
Thus we have:
\be\label{legflsm}
{\partial\over\partial\Lam} \Gam(\p_1,\cdots,\p_n;\Lam)
=\int\!\!{d^4q\over(2\pi)^4}
{q^2{\partial\over \partial\Lam}C_{UV}({q^2}/{\Lam^2})\over 
\left[q^2+C_{IR}(q^2/\Lambda^2)\Sigma(q;\Lam)\right]^2}
E(\q,\p_1,\cdots,\p_n;\Lam),
\ee
where $E(\q,\p_1,\cdots,\p_n;\Lam)$ is as defined in (\ref{legflE})
except now  $G(p;\Lam)$ is defined by
\be\label{Gsm}
G(p;\Lam)\equiv
{C_{IR}(q^2/\Lambda^2)\over
q^2+C_{IR}(q^2/\Lambda^2)\Sigma(q;\Lam)} \nonumber
\ee

Concentrating on the exponential cutoff, $C_{UV}=e^{-q^2/\Lam^2}$,
at one loop we find the renormalized four-point function is given by
\cite{morris2} 
\bea
\lefteqn{\gam(\p_1,\p_2,\p_3,\p_4;\Lam)}
\nonumber \\
& & =  -{\lam}^2\sum_{i=2}^{4}\int_{\Lam}^{\infty}
\!\!d\Lam_1\int\!\!{d^4q\over(2\pi)^4}\left(
{\partial\over{\partial\Lam_1}}{e^{-{q^2}/{{\Lam_1}^2}}\over q^2}
\right)\left\{{\left(1-e^{-{|\q+{\bf{\cal
P}_i}|}^2/{{\Lam_1}^2}}\right)\over
{{|\q+{\bf{\cal P}_i}|}^2}} \right. \label{sm1lp}
\\
& & \hspace{4.0in} \left. -\,
{\left(1-e^{-q^2/{{\Lam_1}^2}}\right)\over{q^2}} \right\}
\nonumber \\
& & = -2{{{\lam}^2}\over (4\pi)^2}\sum_{i=2}^{4}
\int_{\Lam}^{\infty}\!\!{d\Lam_1\over \Lam_1}
\left\{\left({{\Lam_1}\over {\cal P}_i}\right)^2
\left(1-e^{-{{\cal P}_i}^2/2{\Lam_1}^2}\right)-{1\over 2}\right\}
\label{sm1lpex} 
\\
& & = -{{{\lam}^2}\over 2(4\pi)^2}\sum_{i=2}^{4}
\sum_{n=1}^{\infty}{(-1)^n\over (n+1)!\,n}\left({{{\cal P}_i}^2\over
2{\Lam_1}^2}\right)^{n}. \label{sm1lpans}
\eea
The expression in (\ref{sm1lpex}) can be obtained from (\ref{sm1lp})
either by expanding the exponentials, performing the integration over
momentum space and then resumming, or by using
\be
{\left(1-e^{-q^2/{{\Lam}^2}}\right)\over{q^2}}
= {1\over \Lam^2}\int_{0}^{1}\!\!\!da\ e^{-aq^2/\Lam^2}\nonumber
\ee 
and interchanging the order of integration.

We again need to consider the one-loop self energy and once more
calculate its differential to be  
\bea\label{sesmdif}
\lefteqn{{\partial\over\partial\Lam} \Sigma(p;\Lam) = -{{\lam}\over
2}\int\!\!{d^4q\over(2\pi)^4}{1\over q^2}
{\partial\over\partial\Lam}\left(e^{-{q^2}/{\Lam^2}}
\right)}  \nonumber \\
& & \hspace{0.5in}= -{\lam\over (4\pi)^2}\Lam  
\eea
which integrates uniquely to
\be\label{sesm}
\Sigma(p;\Lam)= -{\lam\over (4\pi)^2}{\Lam^2\over 2},
\ee
by virtue of this being a massless theory. While this is exactly the
same value as obtained with a sharp cutoff, it can be shown that this
is a coincidence. To two-loop order, the Legendre flow equation for
a smooth cutoff is  
\bea\label{smmain2lp}
\lefteqn{{\partial\over\partial\Lam}\lam(\Lam) = {1\over \Lam}
{{3{\lam}^2}\over (4\pi)^2}} \nonumber
\\
& & + \, 3{\lam}^3\int\!\!{d^4q\over(2\pi)^4}
{{\partial\over{\partial\Lam}}\left(e^{-{q^2}/{{\Lam}^2}}
\right)\over q^2}\left[{3\Lam^2\over 2(4\pi)^2}
{\left(1-e^{-{q^2}/{\Lam^2}}\right)^2\over q^4} \right. \nonumber
\\
& & -{1\over 2}\int_{\Lam}^{\infty}d\Lam_1
\int\!\!{d^4p\over(2\pi)^4}
{{\partial\over{\partial\Lam_1}}\left(e^{-{p^2}/{{\Lam_1}^2}}\right)
\over p^2}
\left\{3{\left(1-e^{-{p^2}/{\Lam_1^2}}\right)^2\over p^4} \right.
\nonumber
\\
& & {4\left(1-e^{-{|\p+\q|^2}/{\Lam_1^2}}\right)^2\over |\p+\q|^4}
+{8\left(1-e^{-{|\p+\q|^2}/{\Lam_1^2}}\right)
\left(1-e^{-{p^2}/{\Lam_1^2}}\right)\over p^2|\p+\q|^2} \nonumber
\\
& & \left.\left.+{4\left(1-e^{-{q^2}/{\Lam^2}}\right)\over q^2}\left(
{\left(1-e^{-{|\p+\q|}^2/{{\Lam_1}^2}}\right)\over {{|\p+\q|}^2}} 
-{\left(1-e^{-p^2/{\Lam_1^2}}\right)\over p^2}\right)\right\}\right]. 
\eea 

The terms in (\ref{smmain2lp}) are in direct correspondence to those
of (\ref{shmain2lp}). Performing the required integrals we
find the diagram arising from the insertion of the self energy into
the one loop four point function to be $9\ln(4/3){{{\lam}^3}\over
(4\pi)^4}{1\over \Lam}$, while the self energy diagram coming
from the six point vertex with two legs joined is
$-9\ln(4/3){{{\lam}^3}\over (4\pi)^4}{1\over \Lam}$. Again they cancel,
as required.

The next three contributions are all of the form
fig~\ref{figtwo}(b). The first of these gives 
\be\label{sm6pt1}
12{{{\lam}^3}\over (4\pi)^4}{1\over
\Lam}\left(\ln{4\over 3} +
\sum_{n=2}^{\infty}{(-1)^n} 
\left[\,\ln{4\over 3} -{1\over n}
\sum_{s=2}^{n}{n\choose s}{(-1)^s\over s-1}
\left\{1 - {1\over 2^{s-2}} + {1\over
3^{s-1}}\right\} \right]\right),
\ee
(when expanded) which numerically sums to ${{\lam^3}\over (4\pi)^4}{1\over
\Lam}(-2.45411725)$, and the second
\be\label{sm6pt2}
-24{{{\lam}^3}\over (4\pi)^4}{1\over
\Lam}\left(\ln{4\over 3}
+ \sum_{n=1}^{\infty}{(-1)^{n}\over n(n+1)}
\left\{\left({2\over 3}\right)^n -\left({1\over 2}\right)^n  \right\}
\right),
\ee
which sums exactly to $12{{\lam^3}\over (4\pi)^4}{1\over
\Lam} [9\ln3 -2\ln2 -5\ln5]$.
The final part of fig~\ref{figtwo}(b), arising from the iterated
value of $\gam(\p_1,\p_2,\p_3,\p_4;\Lam)$, gives \cite{morris2}
\be\label{smmain}
-12{{{\lam}^3}\over (4\pi)^4}{1\over
\Lam}\sum_{n=1}^{\infty}{{(-1)^{n}}\over 
{n(n+1)}} {1\over 2^n} \left(1 - {1\over
2^{n+1}} \right).
\ee
This can be shown to sum exactly to
$6{{{\lam}^3}\over (4\pi)^4}{1\over \Lam}[6\ln3+4\ln2-5\ln5-1]$.  

For wave function renormalization we have, similarly to before,
\bea
\lefteqn{k^2{\partial\over\partial\Lam}Z(\Lam) =
-{\lam^2}\int\!\!{d^4q\over(2\pi)^4} 
\left({\partial\over\partial\Lam}{e^{-q^2/\Lam^2}\over
q^2}\right)} \nonumber 
\\ 
& & \left. \times \int_{\Lam}^{\infty}d\Lam_1\int\!\!{d^4p\over(2\pi)^4}
\left({\partial\over\partial\Lam}{e^{-p^2/{\Lam_1}^2}\over p^2}\right)
\left({1-e^{-|\p+\q+\k|^2/{\Lam_1}^2}\over |\p+\q+\k|^2}\right)
\right|_{O(k^2)} \nonumber
\\
& & = {{\lam^2}k^2\over (4\pi)^4}{1\over
\Lam}\sum_{n=2}^{\infty}{(-1)^n\over {2^n}},
\eea
which equates to ${1\over 6}{{\lam^2}k^2\over (4\pi)^4}{1\over\Lam}$. 
Hence the $\beta$ function to two loops is given as 
\bea\label{betasm}
\lefteqn{\beta(\Lam) = 3{\lam^2\over (4\pi)^2}
-{12\lam^3\over (4\pi)^4}\left\{\ln{4\over3}
\right.}\nonumber
\\
& & \hspace{0.4in}
+\sum_{n=1}^{\infty}{(-1)^{n}}
\left[\ln{4\over 3}-{1\over12}\left(1\over2\right)^n
+{1\over n(n+1)}
\left(2\left(2\over 3\right)^n -{1\over 2^n}  
- {1\over 2^{2n+1}}\right)
\right. \nonumber
\\
& & \left. \left. \hspace{2in}
-{1\over n+1}\sum_{s=2}^{n+1}{{n+1}\choose s}{(-1)^s\over s-1}
\left\{1 - {1\over 2^{s-2}} + {1\over3^{s-1}}\right\}
\right]  \right\} \nonumber
\\
& & = 3{\lam^2\over (4\pi)^2} -{\lam^3\over (4\pi)^4}
\left[72\ln3-48\ln2-30\ln5+2.45411725+6-{1\over 3}\right],
\eea
which gives the expected form of (\ref{truebeta}).

We see that for smooth exponential cutoff, the derivative expansion series
all have coefficients that fall at least as fast as $1/R^n$, with $R>1$.
(It may
be shown that the $n^{\rm th}$ term in (\ref{sm6pt1}) falls faster than
${1\over n}(2/3)^n$ for large $n$.) It follows then, from the discussion at the
end of the previous section, that the diagrams of fig~\ref{figtwo}
and fig~\ref{figwfn2lp} expanded to the $(2r)^{\rm th}$
power of the external momenta, still
yield convergent derivative expansion series in this case.
 In fact the situation is
even better than this. As before, the $k^{2r}$ term ($k$ some external
momentum) effectively converts the power $q^{2n}$ in the expanded terms
of (\ref{smmain2lp}) to $\sim n^{2r} k^{2r}q^{2n-2r}$, but here the
coefficients of $q^{2n}$ go like $\sim1/n!$ or better and the
power of $q$ is integrated against $e^{-q^2/\Lam^2}$. Converting $q^{2n}$
to $q^{2n-2r}$ means that the $1/n!$ is now incompletely cancelled 
by the $q$ integral, leaving
a remainder $\sim 1/n^{2r}$ for large $n$. Thus the net result is
that not only do the series for all the higher derivative operators converge,
they converge just as fast as the zero external momentum diagrams!

\section{Power law cutoff}

The final cutoff we consider is that of a power law \ie
$C_{UV}(q,\Lam)=1/[1+(q/\Lam)^{2\kappa+2}]$ where $\kappa$ is a
non-negative integer.  Recall that this leads to a well defined
derivative expansion to all orders if $\kappa>D/2-1$ \cite{morris3}
(in the sense that all the momentum integrals converge,
$D$ being the space-time dimension).
Despite this, problems arise with the integrals at two loops,
when we consider perturbation theory. For example, consider
the integral pertaining to fig~\ref{figtwo}(b), obtained by the
iteration of the four-point function at one loop. We have the
following contribution to two-loop $\beta$ function:
\bea\label{pow2lp}
\lefteqn{\sim {\lam^3\over \Lam^{2\kappa +3}} 
\int\!\!{d^4q}{q^{4\kappa}
\over[1+(q/\Lam)^{2\kappa+2}]^3} 
\int_{\Lam}^{\infty}\!\!{d\Lam_1\over \Lam_1^{2\kappa+3}} 
\int\!\!{d^4p}}
\nonumber \\
& & \hspace{0.25in} \times {p^{2\kappa}\over[1+(p/\Lam_1)^{2\kappa+2}]^2}
\left[1- {1\over 1+(|\q+\p|/\Lam_1)^{2\kappa+2}}\right]
{1\over {|\q+\p|^2}}. 
\eea
Derivative expansion requires the inner integral (the one-loop four-point
function) to be expanded in powers of its external momentum $q$. 
The expansion of one-loop terms always
exists to all orders but, once the power $q^{2m}$ is such that
$m\ge\kappa+1$ (or $\kappa+3-D/2$ in general space-time dimension),
the second loop integral over $q$, fails to converge. In this case,
even the {\sl coefficients} of the derivative expansion series are infinite.
In the worst case, $\kappa=2$, this happens at $O(\partial^6)$.

At first sight there is a conflict with our earlier statement, 
that the integrals converge if $\kappa>D/2-1$.
The resolution is that, non-perturbatively there exists also a factor
$\sim 1/[q^2+C_{IR}\Sigma]^3$ from (\ref{legflsm}) and 
(\ref{Gsm}).\footnote{Recall that these equations hold non-perturbatively, 
and it is still the
case that the four-point functions in $E$ need expanding
as a power series in $q$.} For small $\lambda$, $\Sigma\sim \lambda^2q^{2m}$
at $O(\partial^{2m})$, from fig~\ref{figwfn2lp}. The extra powers of $q$
in the denominator always stabilise the integral providing $\kappa>D/2-1$,
but clearly the integral will then diverge as $\lambda\to0$. A little
further analysis taking into account the powers of $\lambda$ in the
four-point functions, shows that in the problematic cases of
$O(\partial^{2m})$ with $m\ge \kappa+3-D/2$,
the contribution behaves overall at small
$\lambda$ as $\sim \lambda^{p}$ where $p=-{2(m-k)+D-4\over m-1}$ 
is a negative fractional power. Obviously then, although the integrals
converge, the results are qualitatively unacceptable.

Clearly these problems are generic, \ie not just for 
contribution fig~\ref{figtwo}(b).

\section{Conclusions and outlook}

We summarise each case in turn.

Using the Wilson/Polchinski equation, the
one-loop $\beta$ function calculation computed via a derivative
expansion, results in a divergent numerical series 
for any cutoffs falling as a power
or as the exponential of any power, in agreement with ref. \cite{morris5}.
Sharp cutoffs require the use of the
Legendre flow equations or equivalent in any case.
However, convergent numerical series are obtained for
certain entire fast falling cutoffs such as
$C_{UV}(x)=\exp\left(1-{\rm e}^x\right)$, or $C_{UV}= \exp\left\{{\rm
e}-\exp\left({\rm e}^x\right)\right\}$.  

For the Legendre flow equations and any cutoff, the one-loop $\beta$
function is exact within the derivative (or momentum-scale) expansion.

Derivative expansions of the Legendre flow equations using power law
cutoffs $\sim 1/x^\kappa$,  fail first at two loops, again as reported
in ref. \cite{morris5}.  In fact here the $O(\partial^{2m})$ result,
where $m\ge\kappa+3-D/2$, diverges. The reason is that certain
contributions at these orders diverge for small coupling $\lambda$, and
thus these $O(\partial^{2m})$ results are simply qualitatively
incorrect in this regime. In $D=4$ dimensions this problem first appears
at $O(\partial^6)$.

The two-loop $\beta$ function computed via a momentum-scale expansion
using sharp cutoff, yields overall a slowly converging numerical
series. The same diagrammatic contributions expanded to $O(k^2)$ in
external momenta $k$ result in series that fail to converge. These
investigations improve on those reported in ref. \cite{morris5} where we
claimed on the basis of incomplete calculations that the two-loop
$\beta$ function yields a fast converging series
\cite{morris1,morris2}. In fact as we have seen, all contributions
converge very fast except for one,
which turns out to be only very slowly converging.  The
analogous contribution to this, at $O(k^2)$, fails to converge at all.

Finally, the two-loop $\beta$ function computed via a derivative
expansion of the Legendre effective action and using an exponential
cutoff $C_{UV}(x)={\rm e}^{-x}$, yields a fast converging numerical
series and moreover, the same diagrams expanded to any power of the
external momenta give series that converge just as fast. These 
results verify and extend those reported in ref. \cite{morris5}.

Clearly we can conclude that sharp cutoffs and power law cutoffs have
inherent limitations, which is rather a shame since these are the only two
cutoffs that continue to preserve reparametrization invariance,
when combined with derivative / momentum-scale expansions
\cite{morris3,morris5}.

On the other hand, use of the Legendre flow equation and exponential
cutoff --precisely
as favoured by a number of authors \cite{ln2,leg2,WS}, yields series
that converge very well to two loops. Although 
our work has been limited at two loops to two-point and four-point
vertices, it is natural to speculate that convergence is found for
all operators at two loops. Certainly it is possible to investigate
other, or maybe all, operators with the techniques used here. Although
we concentrated on four dimensions, it
is clear (from the discussion at the end of sect.5), 
that convergence will be obtained in any dimension $D$.
Naturally the question arises, what happens at three loops?
These calculations are slightly more involved over and above the
extra loop,
because there are now two levels of embedding of subdiagrams, each
needing expansion to a given order in the derivative expansion.
Nevertheless these investigations again look possible and 
the results would be very interesting. 

The use of $C_{UV}(x)={\rm e}^{-x}$ is certainly very helpful in these
investigations since the derivative expansion coefficients can be
calculated exactly, as we have seen.  Clearly the cutoff must
fall faster than a power, but
what else is required to obtain good results?

Although we have identified certain cutoffs to use with the Wilson/Polchinski
equation that give convergent series
for the one-loop $\beta$ function, we did not investigate whether these
series converge for higher derivative operators. This would seem to be
an important and straightforward investigation. Another important
question is the behaviour at
two loops (and higher).

Clearly we leave many questions unanswered. (We have to stop somewhere!)
Nevertheless it must be stressed again that the proof of convergence 
in the many examples we have worked out here, guarantee the accuracy
of high orders of the derivative expansion for these quantities in
the small coupling regime. With the derivative expansion known to be exact
in various other limits (as reviewed in the introduction), these results 
further help to explain the impressive accuracy that may be obtained
with these methods.
 
\section*{Acknowledgements}

TRM acknowledges
support of the SERC/PPARC through an Advanced Fellowship, and PPARC grant
GR/K55738. JFT thanks PPARC for support through a studentship.


\begin{thebibliography}{99}
\bibitem{morris1} T.R. Morris, Int. J. Phys. {  A9} (1994), 2411.
\bibitem{morris4} T.R. Morris, Prog. Theor. Phys. Suppl. {  131}
(1998), 395.
\bibitem{lpa} J.F. Nicoll, T.S. Chang and H.E. Stanley, 
Phys. Rev. Lett. 33 (1974) 540.
\bibitem{largens} F.J. Wegner and A. Houghton, Phys. Rev. A8 (1973) 401;\
M. D'Attanasio and T.R. Morris, Phys. Lett. B409 (1997) 363;\
S.-K. Ma, Rev. Mod. Phys. 45 (1973) 589;\
T.S. Chang, D.D. Vvedensky and J.F. Nicoll, Phys. Rep. 217 (1992) 279;\
K-I Aoki {\it et al}, Prog. Theor. Phys. 95 (1996) 409, hep-ph/9812050.
\bibitem{ln2} M. Reuter, N. Tetradis and C. Wetterich,
Nucl. Phys. B401 (1993) 567.
\bibitem{on} T.R. Morris and M. Turner, Nucl. Phys. B509 (1998) 637.
\bibitem{morris3} T.R. Morris, Phys. Lett. {  B329} (1994), 241.
\bibitem{mass} T.R. Morris, Nucl. Phys. B495 (1997) 477.
\bibitem{morris5} T.R. Morris, Int. J. Mod. Phys. {  B12} (1998), 1343.
\bibitem{morris2} T.R. Morris, Nucl. Phys. {  B458[FS]} (1996), 477.
\bibitem{zak} T.R. Morris, in {\it New Developments in Quantum Field Theory},
NATO ASI series B, 366, Plenum Press 1998, hep-th/9709100.
\bibitem{shaf} J.D. Shafer and J.R. Shepard, Phys. Rev. D51 (1995) 7017;
D55 (1997) 4990.
\bibitem{Ball} R.D. Ball, P.E. Haagensen, J.I. Latorre, and E. Moreno,
Phys. Lett. {  B347} (1995), 80.
\bibitem{threeps} G.R. Golner, Phys. Rev. {  B33} (1986), 7863;\
 J. Comellas, Nucl. Phys. {  B509} (1998), 662.
\bibitem{polder} G. Zumbach, Nucl. Phys. B413 (1994) 754, 
Phys. Lett. A190 (1994) 225;\
K.-I. Aoki {\it et al}, Prog. Theor. Phys. (1996) 409,
hep-th/9812050;\
J. Comellas and A. Travesset, 
Nucl. Phys. {  B498} (1997), 539;\
Y. Kubyshin, Int. J. Mod. Phys. {  B12} (1998), 1321;\ 
R. Neves, Y. Kubyshin and R. Potting, hep-th/9811151.
\bibitem{twod} T.R. Morris, Phys. Lett. B345 (1995) 139.
\bibitem{leg} J.F. Nicoll and T.S. Chang, Phys. Lett. 62A (1977) 287;\
M. Bonini {\it et al}, Nucl. Phys. B409 (1993) 441.
\bibitem{leg2}C. Wetterich, Phys. Lett. B301 (1993) 90.
\bibitem{Christoph} C. Wetterich, private communication.
\bibitem{WS} N. Tetradis and C. Wetterich, Nucl. Phys. {  B422} (1994), 541;
B398 (1993), 659;\
J. Berges, N. Tetradis and C. Wetterich, Phys. Lett. B393 (1997) 387,
Phys. Rev. Lett. 77 (1996) 873;\
D.U. Jungnickel and C. Wetterich, hep-ph/9902316;\
S. Seide and C. Wetterich, cond-mat/9806372;\
J. Berges and C. Wetterich, Nucl. Phys. B487 (1997) 675;\
S. Bornholdt {\it et al}, Int. J. Mod. Phys. A14 (1999) 899,
Phys. Rev. D53 (1996) 4552, Phys. Lett. B348 (1995) 89;\ 
M. Grater and C. Wetterich, Phys. Rev. Lett.
75 (1995) 378;\ D. Litim {\it et al}, Mod. Phys. Lett. A12 (1997) 2287;\
N. Tetradis, Phys. Lett. B409 (1997) 355.
\bibitem{bonini} M. Bonini {\it et al}, Nucl. Phys. {  B483} (1997) 475.
\bibitem{wilson} K. Wilson and J. Kogut, Phys. Rep {  12C} (1974), 75. 
\bibitem{pol} J. Polchinski, Nucl. Phys. {  B231} (1984), 269.
\bibitem{zinn} \eg J. Zinn-Justin {\it Quantum Field Theory and
Critical Phenomena} (C.P., Oxford, 1993). 
\end{thebibliography}
\end{document}